\documentclass[iop]{emulateapj}
\usepackage{natbib}
\usepackage[breaklinks,colorlinks,urlcolor=blue,citecolor=blue,linkcolor=blue]{hyperref} 
\submitted{Accepted to AJ} 

\newcommand{\kms}{km~s$^{-1}$}
\newcommand\lbol{\ensuremath{L_{\text{bol}}}}
\newcommand\slr{\ensuremath{L_{\nu}}}
\newcommand\slumcgs{erg~s$^{-1}$~Hz$^{-1}$}
\newcommand\ujy{$\mu$Jy}
\newcommand\vsi{\ensuremath{v \sin i}}

\begin{document}

\title{WISEP J060738.65+242953.4: A Nearby. Pole-On L8 Brown Dwarf with Radio Emission}

\author{John E.\ Gizis}
\affil{Department of Physics and Astronomy, University of Delaware, Newark, DE 19716, USA}
\author{Peter K.\ G.\ Williams}
\affil{Harvard-Smithsonian Center for Astrophysics, 60 Garden Street, Cambridge, MA 02138, USA}
\author{Adam J.\ Burgasser}
\affil{Center for Astrophysics and Space Science, University of California San Diego, La Jolla, CA 92093, USA} 
\author{Mattia Libralato\altaffilmark{1}, Domenico  Nardiello\altaffilmark{1}, Giampaolo Piotto\altaffilmark{1}}
\affil{Dipartimento\ di Fisica e Astronomia, Universit\`a di Padova, Vicolo dell'Osservatorio 3, Padova, I-35122, Italy}
\author{Luigi  R.\ Bedin}
\affil{INAF-Osservatorio Astronomico di Padova, Vicolo dell'Osservatorio 5, Padova, I-35122, Italy}
\author{Edo Berger}
\affil{Harvard-Smithsonian Center for Astrophysics, 60 Garden Street, Cambridge, MA 02138, USA}
\author{Rishi Paudel}
\affil{Department of Physics and Astronomy, University of Delaware, Newark, DE 19716, USA}
\altaffiltext{1}{INAF-Osservatorio Astronomico di Padova, Vicolo dell'Osservatorio 5, Padova, I-35122, Italy}

\begin{abstract}
We present a simultaneous, multi-wavelength campaign targeting the nearby (7.2 pc) L8/L9 (optical/near-infrared) dwarf WISEP J060738.65+242953.4 in the mid-infrared, radio, and optical. Spitzer Space Telescope observations show no variability at the 0.2\% level over 10 hours each in the 3.6 and 4.5 micron bands. {\it Kepler} K2 monitoring over 36 days in Campaign 0 rules out stable periodic signals in the optical with amplitudes great than 1.5\% and periods between 1.5 hours and 2 days.  Non-simultaneous Gemini optical spectroscopy detects lithium, constraining this L dwarf to be less than $\sim 2$ Gyr old, but no Balmer emission is observed.  The low measured projected rotation velocity ($v \sin i < 6$ km s$^{-1}$) and lack of variability are very unusual compared to other brown dwarfs, and we argue that this substellar object is likely viewed pole-on. 
We detect quiescent (non-bursting) radio emission with the VLA. Amongst radio detected L and T dwarfs, it has the lowest observed $L_\nu$ and the lowest $v \sin i$.
We discuss the implications of a pole-on detection for various proposed radio emission scenarios.
\end{abstract}


\section{Introduction\label{intro}}

Brown dwarfs are doomed to steadily cool and fade by their lack of hydrogen fusion, but they nevertheless exhibit a wide variety of non-equilibrium, time-dependent behaviors. Their mineral and metal condensate clouds are not completely uniform, resulting in periodic variability and ``weather" \citep{2014ApJ...782...77B,2015ApJ...799..154M}.  The dramatic differences in observed spectra at the L/T transition are typically attributed to 
changes in the qualitative properties of clouds \citep{1999ApJ...520L.119T,Burrows:2000rt,2001ApJ...556..357A,Ackerman:2001fj,Burgasser:2002lr,2004AJ....127.3553K}. In some cases, rapid, high-amplitude variability is observed \citep{2014ApJ...793...75R}. These cloud changes are so significant that they may even affect the luminosity evolution of the brown dwarf itself \citep{2008ApJ...689.1327S,2015ApJ...805...56D}. Meanwhile, magnetic fields are generated even in L and T dwarfs, resulting in surprisingly strong, variable radio emission \citep{Berger:2002fk,McLean:2012qy,2015ApJ...808..189W,2016ApJ...818...24K}.  

Deeper understanding of these phenomena requires both surveys of typical ultracool (later than M7) dwarfs and detailed studies of particularly favorable targets.  One such target is the nearby brown dwarf WISEP J060738.65+242953.4 (\citealt{Castro:2011W}, hereafter W0607+24). Classified as L8 in the optical and L9 in the near-infrared, with a preliminary trigonometric parallax placing it at a distance of $7.19^{+0.11}_{-0.10}$ pc \citep{2013ApJ...776..126C}, W0607+24 is the nearest known late-L dwarf northern hemisphere, and the third-nearest on the whole sky. W0607+24 is therefore a prime target in understanding the physics of the L/T transition, across which mineral condensate clouds are believed to sink below the photosphere (see \citealt{2016ApJ...817L..19T} for an alternate, cloudless model).  W0607+24 also lies close to the ecliptic plane, in the K2 mission \citep{2014PASP..126..398H} Campaign 0 field.
Although designed to search for transiting planets around bright Sun-like stars, the {\it Kepler} space telescope \citep{2010ApJ...713L..79K} can also obtain long time-series photometry of fainter targets that happen to lie in the field of view. Indeed, during its original mission,  it was able to measure rotational modulations due to photospheric spots in late-M \citep{2013A&A...555A.108M} and L1 \citep{2013ApJ...779..172G} very-low-mass stars. Each K2 field offers the opportunity to monitor additional very-low-mass stars or brown dwarfs. K2 Campaign 2 monitoring of Upper Scorpius has detected variability in 16 young, M-type brown dwarfs \citep{2015ApJ...809L..29S}.
Motivated by the K2 observations of W0607+24, we observed it simultaneously with the Karl G. Jansky Very Large Array (VLA)\footnote{The VLA is operated by the National Radio Astronomy Observatory, a facility of the National Science Foundation operated under cooperative agreement by Associated Universities, Inc.} and the Spitzer Space Telescope  \citep{2004ApJS..154....1W}. In this paper, we present the results of ground-based spectroscopy and the simultaneous multi-wavelength observations.

\section{Data and Observations\label{sec-data}}

\subsection{Spectroscopy}

\begin{figure*}
\plotone{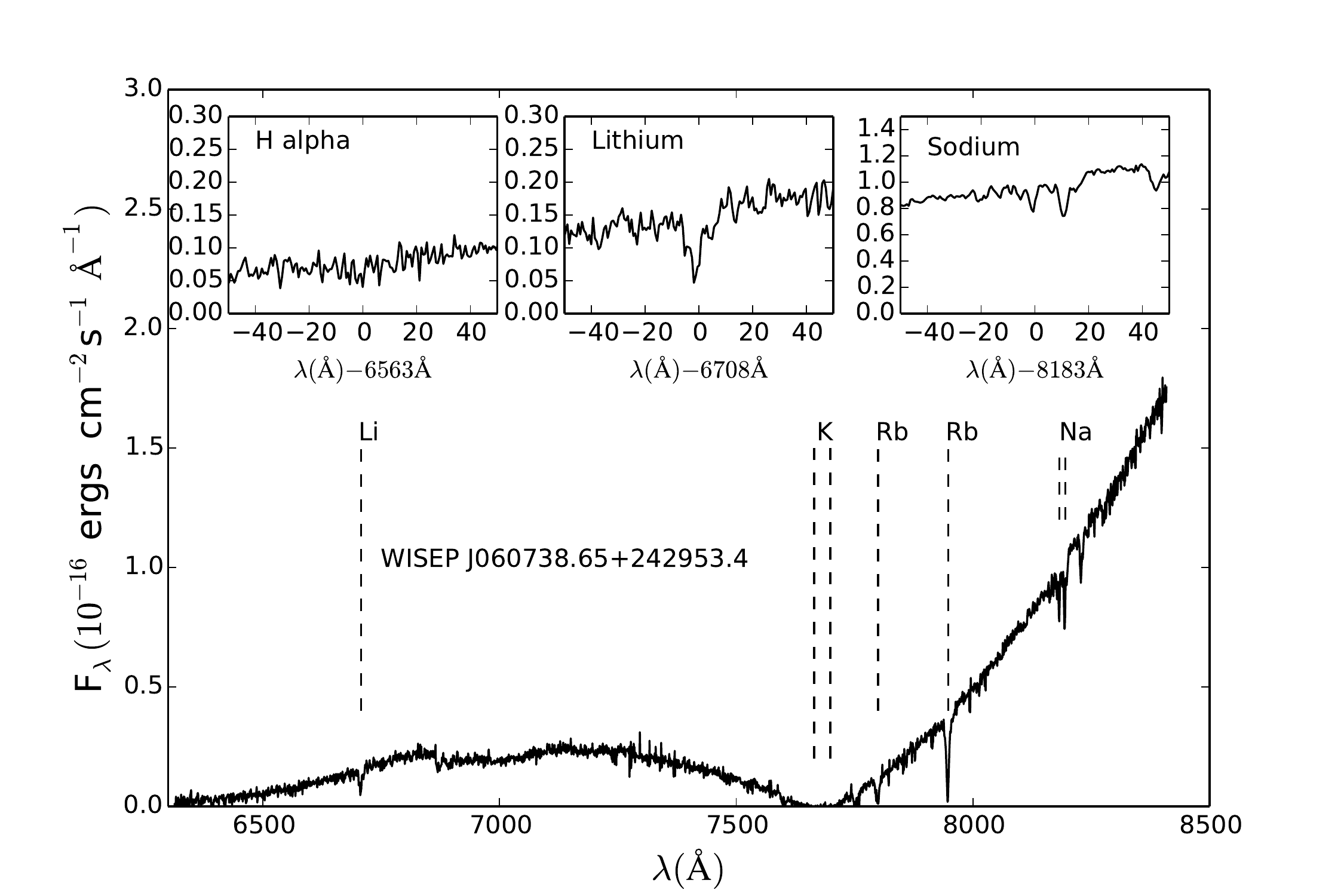}
\caption{Optical spectrum of W0607+24.  We detect lithium absorption but no H$\alpha$ emission or absorption. \label{fig-optical}}
\end{figure*}

We observed W0607+24 on UT Date 2012 October 15 with the Gemini-North telescope (Gemini program GN-2012B-Q-105) and the GMOS spectrograph \citep{Hook:2004lr} using grating R831. The wavelength coverage was 6340 to 8460\AA~with a resolution of $\sim2$\AA. We processed it using standard IRAF Gemini tasks and co-added the four 600 second exposures. Conditions were non-photometric. The spectrum (Figure~\ref{fig-optical}) is consistent with the L8 optical spectral type previously reported \citep{2013ApJ...776..126C}. We detect lithium absorption (equivalent width EW $4.0 \pm 0.4$ \AA, Table~\ref{tab1}). There is no detectable H$\alpha$ emission or absorption (EW $<0.5$\AA).  

\begin{figure*}
\includegraphics[width=\linewidth]{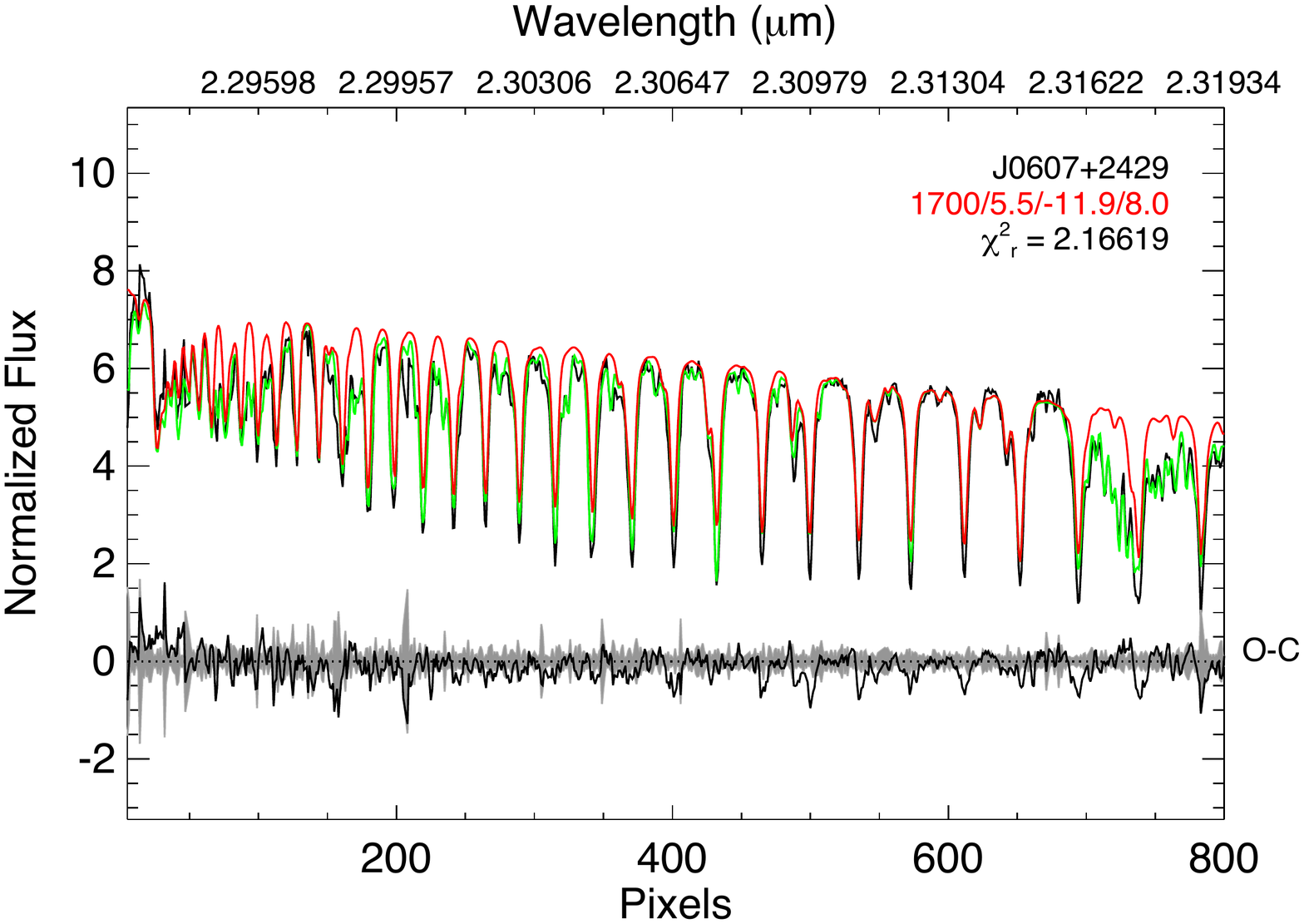}
\caption{NIRSPEC spectrum of W0607+24 (black line) with a model without (red) and with (green) telluric absorption. Pixel scale is listed along the bottom while wavelength scale is listed along the top. The difference between data and model (O-C) is shown in black at the bottom of the plot; the $\pm 1 \sigma$ uncertainty spectrum is shown in grey. \label{fig-keck}}
\end{figure*}

We observed W0607+24 on UT date 2013 October 16 with the Keck II NIRSPEC near-infrared echelle spectrograph \citep{McLean:2000lr} in partly cloudy conditions and excellent seeing (0$\farcs$5) at 2~$\micron$. The high-dispersion mode was used with the 0$\farcs$432$\times$12$\arcsec$ slit and N7 filter to obtain 2.00--2.39~$\mu$m spectra over orders 32--38 with $\lambda/\Delta\lambda$ = 20,000 ($\Delta{v}$ = 15~km~s$^{-1}$) and dispersion of 0.315~{\AA}~pixel$^{-1}$. Two 900~s exposures were obtained at an airmass of 1.03 in two nods separated by 7$\arcsec$ along the slit, and the nearby A0~V star HD~43584 ($V$ = 5.11) was observed for telluric and flux calibration. Raw data frames were reduced and combined following standard procedures, and spectra in order 33 (2.293--2.320~$\micron$) were optimally extracted. The 1D uncalibrated spectrum was then fit to a suite of spectral templates constructed from the BT-Settl atmosphere models \citep{Allard:2011uq,2014IAUS..299..271A}, telluric absorption modeled from the Solar atlas of \citet{Livingston:1991fj}, and a 2nd-order polynomial continuum, using a Markov-Chain Monte Carlo (MCMC) code with Metropolis-Hastings algorithm; see \citet{2015AJ....149..104B} for details. We compared solar-metallicity models in the $T_{\rm eff}$ = 1300--2100~K and $\log{g}$ = 4.0--5.5 (cgs) ranges.  Both wavelength scale, source $v_{\rm rad}$ and $v\sin{i}$ rotational broadening were allowed to vary to minimize $\chi^2$ residuals.  The spectrum in the MCMC simulation with the lowest residuals is shown in Figure~\ref{fig-keck} to illustrate the quality of the fits. The outcome of the MCMC simulation is the posterior probability density function for the five-dimensional parameter space.  By marginalizing (integrating) over the other parameters, we find mean values of $v_{\rm rad}$ = $-$11.9$\pm$1.1~km~s$^{-1}$ and $v\sin{i}$ = 6$\pm$2~km~s$^{-1}$.  However, the last value is suspect, as the broadening profile is actually smaller than the instrumental profile.  W0607+24 has the narrowest lines observed so far in the ongoing survey of \citet{2012ApJ...757..110B}, so we conservatively claim a limit on the rotational broadening of $v\sin{i} <$ 6~km~s$^{-1}$.

\subsection{Radio Observations and Analysis}

We observed W0607+24 with the VLA for 1.5~hr starting at UT date 2014 May 23
22:04 (BMJD 56800.9298; program 14A-541) using two frequency windows of 1~GHz
bandwidth each, centered at 5.0 and 7.1~GHz. The gain and phase calibrator was
J0559$+$2353 and the bandpass and flux density calibrator was
\object{3C\,138}. We calibrated the data using standard techniques in the CASA
software environment \citep{the.casa}, using the ``Perley-Butler 2010'' flux
density scale \citep{2013ApJS..204...19P} and flagging RFI automatically using the
\textsf{aoflagger} tool \citep{2010MNRAS.405..155O, 2012A&A...539A..95O}. After imaging the calibrated
visibilities, we detect a 3.5$\sigma$ source 0.6$''$ distant from the
predicted position of W0607+24 having a total flux density of $15.6 \pm
6.3$~\ujy\ at a mean frequency of 6.05~GHz, where the flux density uncertainty
is determined from non-linear least squares modeling of the image data.
Imaging the two spectral windows separately, we measure flux densities of
$16.5 \pm 7.6$ and $15.6 \pm 10.7$~\ujy\ at 5.0 and 7.1~GHz, respectively,
corresponding to a limit on the spectral index $|\alpha| \lesssim 2$. We
imaged the Stokes~$V$ data, finding no evidence for circularly polarized
emission from W0607+24 in an image with rms noise of 4.5~\ujy, leading to a
3$\sigma$ limit on the fractional circular polarization of $|V|/I \lesssim
90$\%. We extracted a radio light curve for W0607+24 using the technique
described in \citet{2013ApJ...767L..30W} and found no evidence of variability.

\subsection{Mid-Infrared Photometry \label{sec:midir}}

\begin{figure*}
\plotone{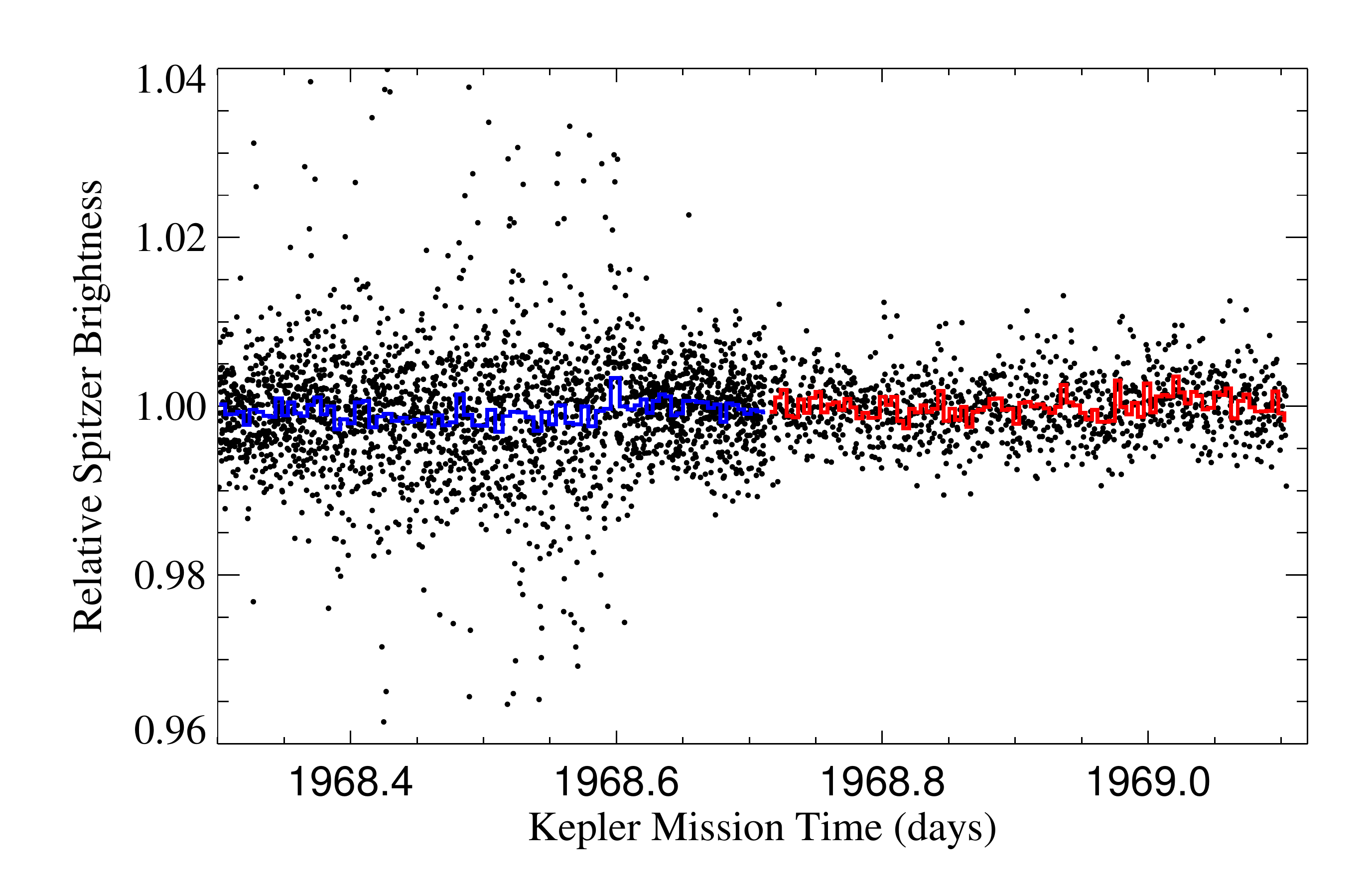}
\caption{Spitzer photometry of W0607+24. The individual frame measurements are shown as points and the averaged photometry (blue for [3.6], red for [4.5]) is shown as a histogram.  There is no evidence of variability in this brown dwarf in either band. The time series begins at BMJD 56800.8010689. \label{fig-spitzer}}
\end{figure*}

We observed W0607+24 with the IRAC camera \citep{2004ApJS..154...10F} under Spitzer DDT program 10167 for twenty hours on UT Dates 2014 May 23-24. 
The principal limit to high precision relative photometry with IRAC is intra-pixel sensitivity variations coupled to telescope pointing changes \citep{2012SPIE.8448E..1IG,2012SPIE.8442E..1YI}.  We designed the observations using the procedures developed by the Spitzer Science Center (SSC). W0607+24 was placed on the IRAC ``sweet spot" (pixels $X=23$, $Y=231$). Following a thirty minute staring observation to allow the telescope to settle, we observed with IRAC Channel 1 (hereafter [3.6]) for 2709 twelve-second exposures starting at BMJD 56800.8010689 (after discarding the first frame). We then re-positioned the telescope and observed with IRAC channel 2 (hereafter [4.5]) for 1082 thirty-second exposures starting at BMJD 56801.2149001. We measured the target centroid using the box centroider software provided by SSC and then measured aperture photometry with a radius of 3 pixels. We analyzed the photometry following the techniques described by \citet{Heinze:2013uq}.  First, outliers are rejected by a robust fitting algorithm. We then median smoothed the centroids on a four-to-five minute timescale (combining 25 observations for [3.6] and 7 for [4.5]), a period much shorter than the instrumental or astrophysical timescales. We find that the telescope pointing has a 53 minute periodic component with an amplitude of 0.02 pixels in $X$ and $Y$; additionally, over ten hours there is a drift of 0.05 pixels in $X$ and 0.06 pixels in $Y$.  We then modeled the pixel phase photometric shift in each band with a linear function in $X$ and $Y$, which removed the correlation of the measured photometry with the telescope pointing.  
In Figure~\ref{fig-spitzer}, we plot the measured relative photometry, where we have normalized each band by its median value. The histogram shows the data averaged over 30 measurements for [3.6] and 12 points for [4.5].  (The individual [3.6] points are noisier due to the shorter exposure time.)  We find that the brightness of W0607+24 is constant in time for both bands (Figure~\ref{fig-spitzer}). The standard deviation of the averaged [3.6] data is 0.12\% and the [4.5] data is 0.13\%. 

To convert our 3-pixel aperture photometry to calibrated Vega-system magnitudes on the \citet{2005PASP..117..978R} system, we use the SSC aperture corrections and the SSC array location-dependent photometric correction to find that W0607+24 has $m_{[3.6]} = 11.13$ and $m_{[4.5]} = 10.95$. The uncertainties in these magnitudes are dominated by systematic uncertainties in the corrections, $\sim \pm 0.03$ mag. The observed color of 0.18 is similar to other L8 dwarfs measured by \citet{2006ApJ...651..502P}.  

We do not observe any significant evidence of variability, whether due to rotation or ``weather."
To quantify the variability limits, we use a Markov Chain Monte Carlo simulation to generate sine function fits to each of the bands.  Each sine curve has the form  $a_0 + a_1 \sin (t/P + \phi)$) with periods $0<P<1$ days for Spitzer. After an initial burn-in of 2000 steps, we run one million steps.  For Spitzer [3.6], 99\% of the fits have amplitude $a_1 < 0.0019$. Similarly, for Spitzer [4.5],  99\% of the fits have $a_1 < 0.0017$. This rules out rotational variability at the 
levels ($>0.2 \%$) seen in most L dwarfs \citep{2015ApJ...799..154M}.

\begin{deluxetable}{lcl}
\tablewidth{0pc}
\tabletypesize{\footnotesize}
\tablecaption{WISEP J060738.65+242953.4}
\tablehead{
\colhead{Parameter} & 
\colhead{W0607+24} & 
\colhead{Remarks}}
\startdata
Sp. Type (nIR) & L9 &  \citet{2013ApJ...776..126C} \\
Sp. Type (opt) & L8 &  \citet{2013ApJ...776..126C} \\
$\pi_{abs}$ & $138 \pm 2$ & mas, \citet{2013ApJ...776..126C} \\
$v_{\rm rad}$ & $-11.9 \pm 1.1$ & \kms \\
$v \sin i $ & $<6$ & \kms \\
EW Lithium & $4.0 \pm 0.4$ &\AA \\
EW H$\alpha$ & $<0.5$ & \AA \\
$\log \slr$  & $12.0\pm0.2$ & \slumcgs \\ 
$\log \slr / \lbol $ &  ${-16.9 \pm 0.2}$ & Hz$^{-1}$ \\
U & $+12.9 \pm 1.3$ & \kms \\
V & $-0.7 \pm 0.6$ &\kms\\
W & $-20.0 \pm 0.5$ & \kms\\ 
$K_P$ & $19.73  \pm 0.08 $& mag \\
$m_{[3.6]}$ & $11.13 \pm 0.03$ & mag \\
$m_{[4.5]}$ & $10.95 \pm 0.03$ & mag \\
$m_{bol} $ & $15.70 \pm 0.03$ & mag \\
$\log L/L_\odot$ & $-4.66 \pm 0.02$ \\
$\log L_{{\rm H}\alpha}/L_{\rm bol}$ & $<3\times10^{-7}$ \\
Age & $<2$ & Gyr \\
Mass & $\le 0.055$ & $M_\odot$ 
\enddata
\label{tab1}
\end{deluxetable}

\subsection{Optical Photometry}

\begin{figure}
\plotone{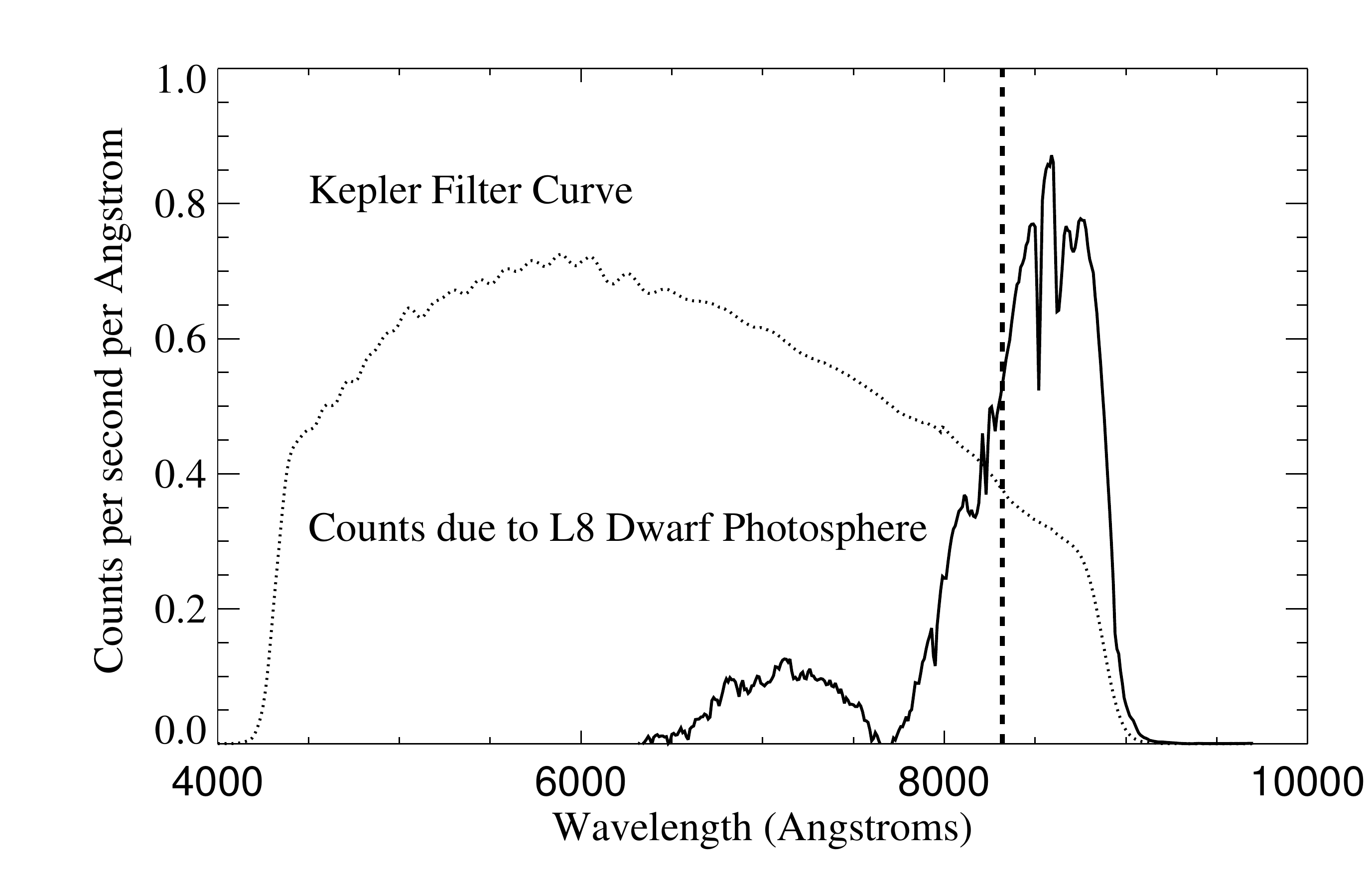}
\caption{The predicted counts as a function of wavelength for W0607+24 (solid curve).  The optical spectrum of \citet{2013ApJ...776..126C} has been folded through the {\it Kepler} filter curve (dotted). The effective wavelength of 8320\AA~for this target is shown as a dashed line.   \label{fig-keplerfilter}}
\end{figure}

\begin{figure*}
\epsscale{1.3}
\plotone{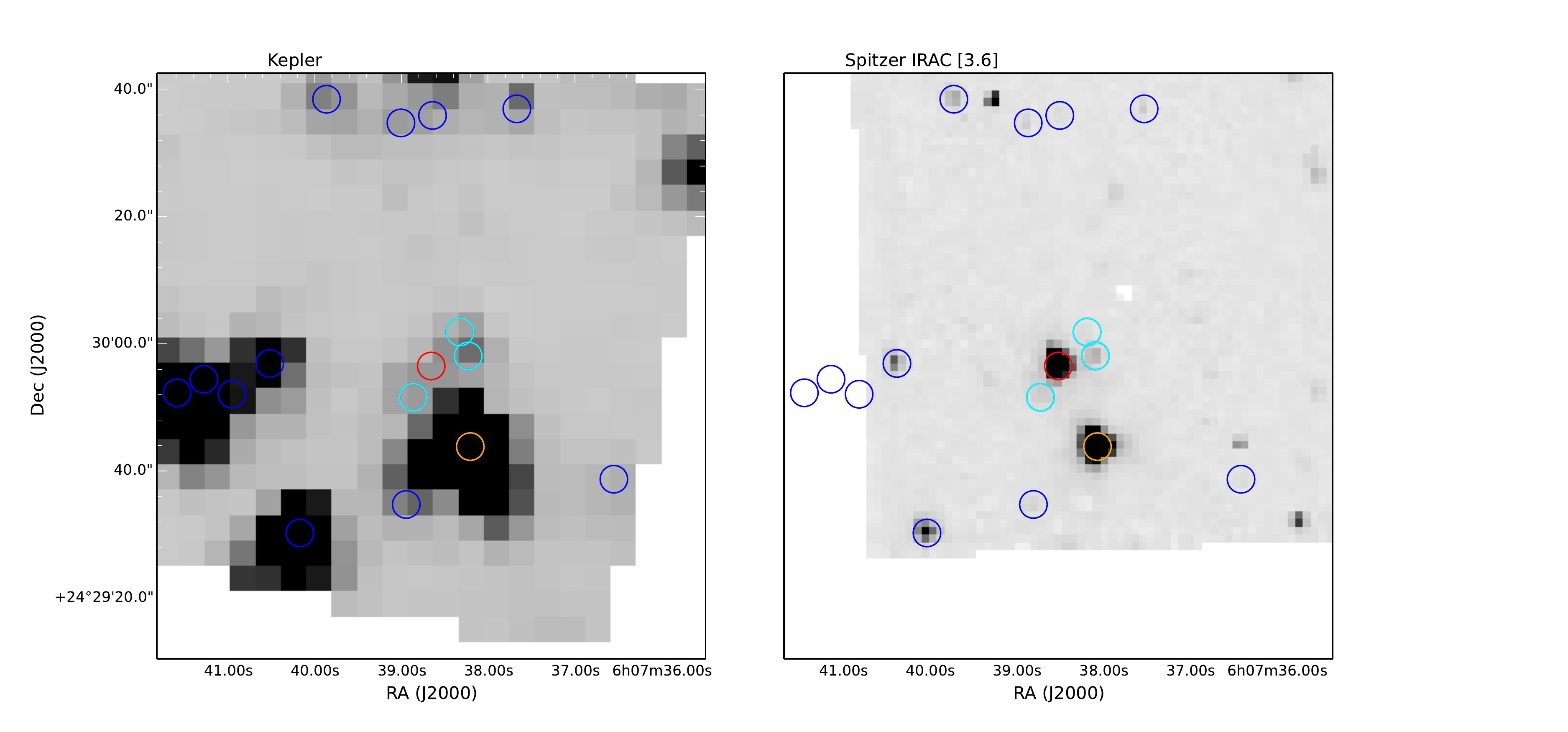}
\caption{K2 and Spitzer [3.6] images of W0607+24. The position of W0607+24 is marked by a red circle. SDSS J060738.06+242939.5, marked by an orange circle, is the brightest star in the K2 image. The three faint ($r =18.3-19.6$) stars marked by cyan circles, contribute significant flux to the K2 measurement but are negligible at Spitzer's longer wavelength and higher resolution. Other stars with $r<20$ in SDSS are marked with blue circles. \label{fig-fov}}
\end{figure*}

W0607+24 was monitored in K2 Campaign 0 as source 202059521 under program GO0008.  For each 30-minute long cadence exposure \citep{Jenkins:2010fk}, a 22 by 21 pixel aperture was saved. Each pixel is 4 by 4 arcseconds. 
Folding the observed W0607+24 optical spectrum from \citet{2013ApJ...776..126C} through the {\it Kepler} spectral response curve \citep{2010ApJ...713L..79K} and scaling from the observed {\it Kepler} count rate (550 DN s$^{-1}$) for the L1 dwarf WISEP J190648.47+401106.8 (\citealt{2013ApJ...779..172G}, W1906+40), we expect a K2  count rate of $\sim 100$ DN s$^{-1}$.
As shown in Figure~\ref{fig-keplerfilter}, it is the red tail of the Kepler filter that allows a detection; 
we calculate that the effective wavelength of the Kepler filter is 832~nm for W0607+24. However,
crowding is important at W0607+24's low galactic latitude (Figure~\ref{fig-fov}).  
The brightest star in the Kepler image, SDSS J060738.06+242939.5 ($r=13.66$), is only 16 arcseconds away. In addition there are three fainter stars between 6 and 8 arc seconds away from W0607+24: SDSS J060738.72+242947.2 ($r=18.67$), SDSS J060738.18+242957.5 ($r=19.83$), and SDSS J060738.09+242953.8 ($r=18.26$). 

\begin{figure}
\epsscale{1.3}
\plotone{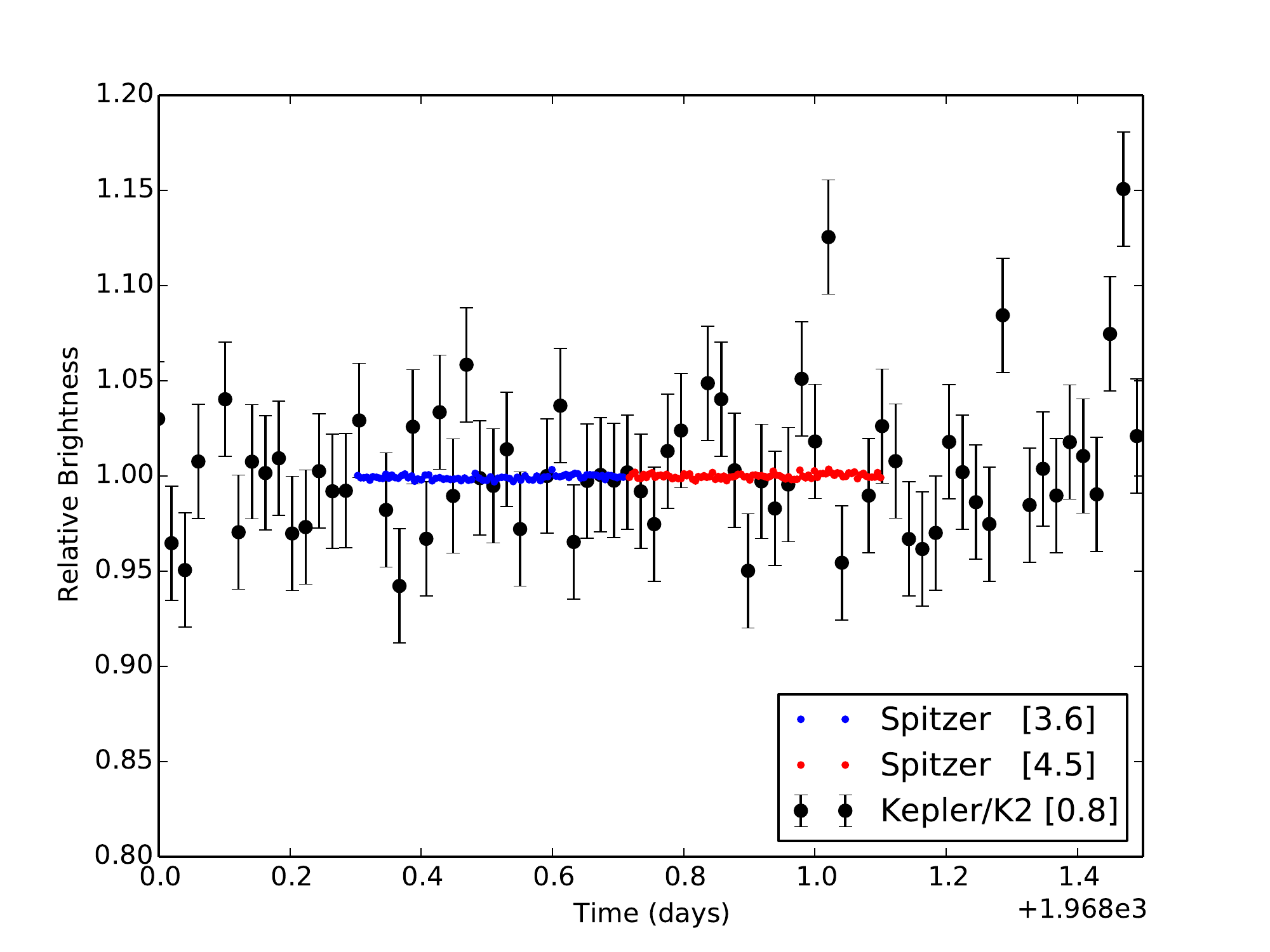}
\caption{  {\it K2} photometry compared with Spitzer photometry. The {\it K2} data are consistent with a constant light curve with an uncertainty (standard deviation) of 3.7\%. \label{fig-compare3}}
\end{figure}

\begin{figure}
\epsscale{1.3}
\plotone{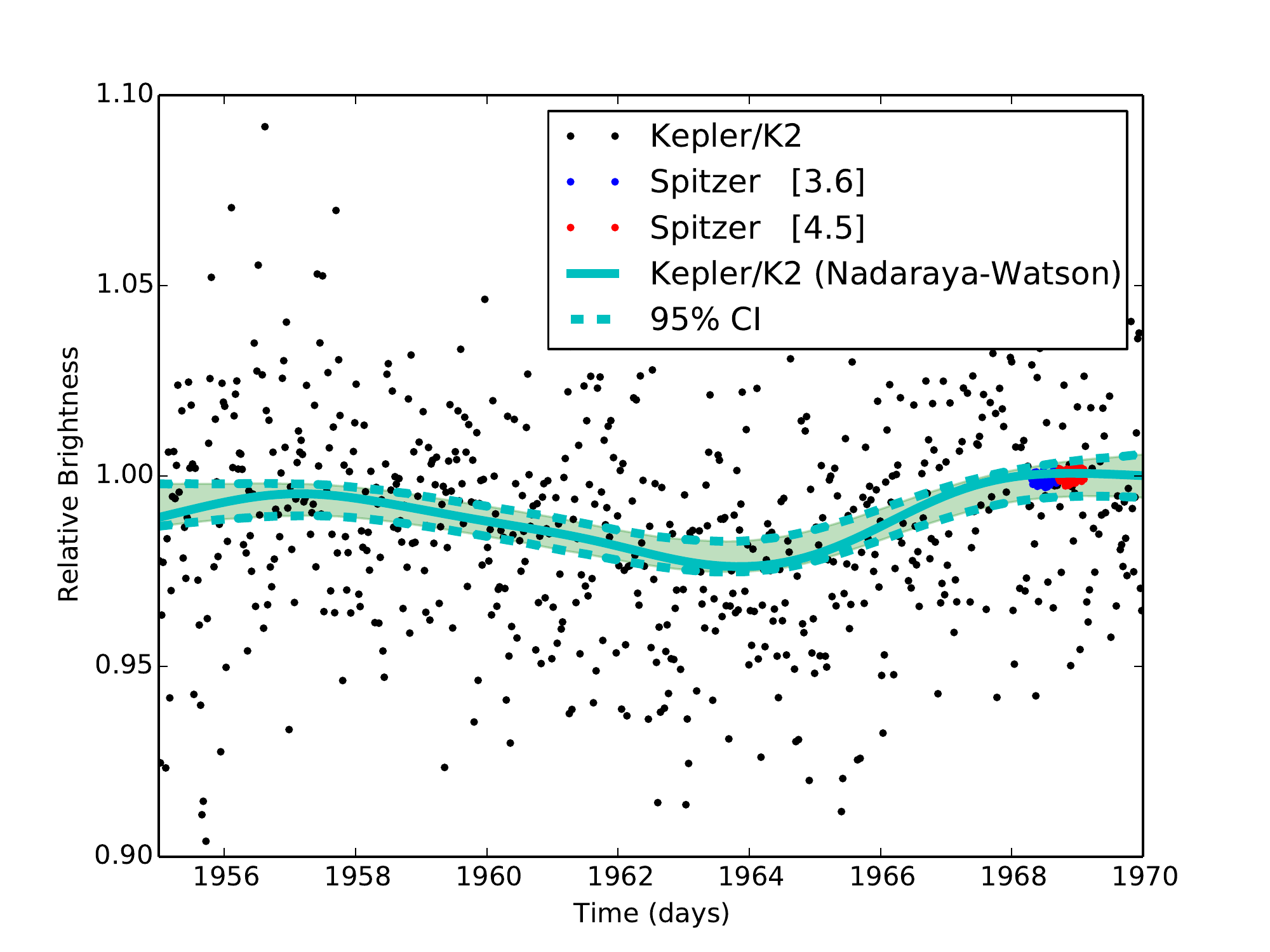}
\caption{K2 photometry of W0607+24 for a 15 day period with the most stable spacecraft pointing. The non-parametric Nadaraya-Watson kernel regression of the trend is shown. We find no evidence of a long rotation period ($P>10$ hr).  The long-term variations are most likely to due to systematic instrumental (pointing) effects. \label{fig-k2photall}}
\end{figure}

\citet{2016MNRAS.456.1137L} have demonstrated that effective pixel-convolved
PSF (ePSF, \citealt{2000PASP..112.1360A}) neighbor-subtraction with a high angular resolution input star list allows photometry of even faint stars in the K2 Campaign 0 observations of open cluster M\,35 and NGC\,2158.  W0607+24 is on the same K2 detector as M\,35, so we use the same ePSF models and methodology.  The input star catalog comes from the high-angular-resolution input star list from the Asiago Schmidt telescope used in the M\,35 analysis ({\it Asiago Input Catalog}, \citealt{2015MNRAS.447.3536N}), but W0607+24's position is updated with an early 2015 image. After subtracting the neighbor stars, we measured both ePSF-fit and aperture (1 pixel) photometry for W0607+24; these are consistent, so we base the remainder of our analysis on the ePSF photometry results. We use only the data from the final 36 days of Campaign 0 ({\it Kepler} mission times 1935.9 to 1972.2) during which the highest-quality data were collected. We discard all photometry during thruster fires and de-trend the drift-induced effects as in \citet{2016MNRAS.456.1137L} using the data for all stars in the M35 region; this removes the spurious instrumental periodic 5.9-hour signal.  (See \citet{2014PASP..126..948V} for a discussion of the effects of drift on the photometry.)  Applying the \citet{2016MNRAS.456.1137L} instrumental calibration the median observed K2 magnitude of W0607+24 is $K_P = 19.72$.  Similar magnitude M35 stars have point-to-point root-mean-square of $\sim 3\%$. 
The K2 photometry during the time period of the Spitzer observations is shown in Figure~\ref{fig-compare3}.  It is consistent with a constant light curve and the expected noise.  
Over the full 36 day period, the photometry drifts by 0.12 magnitude, mainly before day 1955. This drift is correlated with the systematic change of the positions of the stars on the detectors and we therefore attribute it to the instrumental effects. The {\it Kepler} median count rates of \citet{2015ApJ...813..104G} for W1906+40 varied by $\pm 0.08$ mags for different spacecraft quarters, so we adopt this as the uncertainty for the W0607+24 photometry. Searching both the entire dataset, ten day subsets, and five day subsets, there are no significant periodogram peaks for periods $<2$ days in the K2 data. The light curve after day 1955 is shown in Figure~\ref{fig-k2photall}.  We show a non-parametric Nadaraya-Watson kernel regression (computed using the PyQt package; see \citealt{2012msma.book.....F} for a discussion of this statistical method); the $\sim2$\% variations  on week-long timescales can be attributed to the K2 pointing drift.  Simulations similar to those of Section~\ref{sec:midir} show stable sinusoidal signals with periods less than 2 days and amplitudes $>1.4\%$ can be ruled out.

\section{Discussion}

\subsection{The orientation of W0607+24\label{orientation}}

The spectroscopy of W0607+24 indicates it is a typical L8-L9 dwarf. Integrating the observed spectra and photometry, we find that the luminosity is $\log L_{\rm bol}/L_\odot = -4.66 \pm 0.02$. This luminosity and the detection of lithium together imply it is at or below the lithium-burning limit ($M \le 0.055 M_\odot$) with an age less than 2 billion years according to evolutionary models \citep{1997ApJ...491..856B,2003A&A...402..701B,2008ApJ...689.1327S}. These models also predict the radius, $R \approx 0.1 R_\odot$.  

The observed projected rotation period and the model radius imply a rotation period of $P>20$ hours if $i=90^\circ$, or $P>16$ hours for the expected stellar average $\sin i = \pi/4$. Although we cannot directly rule out such a slow rotation, it is in conflict with our prior expectations from other studies of brown dwarfs.  Very few 
brown dwarfs have low $v \sin i$: \citet{2008ApJ...684.1390R} found 1 of 45 L dwarfs have $v \sin i < 3$~\kms; \citet{Blake:2010gf} found 4 of 57 L dwarfs have $v \sin i < 9$~\kms; \citet{2006ApJ...647.1405Z} had 1 of 15 L and T dwarfs with $v \sin i < 15$~\kms. \citet{2014ApJ...793...75R} argue that these collective $v \sin i$~ measurements are consistent with a single log-normal distribution peaked at $P=4.1$ hrs with $\sigma_{\ln P}=0.48$. 
If we simulate ten million randomly inclined objects\footnote{If $x$ is a randomly generated real number between 0 and 1 with a uniform distribution, then $\cos i = 2x-1$.}  with radius $R=0.1R_\odot$ drawn from this log-normal period distribution, we find that 3.4\% have $v \sin i <6$~\kms.  The distribution of true periods and inclinations for the simulated objects with $v \sin i <6$ \kms~is given in Table~\ref{simulation}.  This distribution corresponds to the Bayesian posterior probabilities given the prior period distribution and the evidence of the observed $v \sin i$. There is only a 5\% chance of a period of 15 hours or greater; on the other hand, there is a 77\% chance that $i<20^\circ $. The median rotation period for these objects is 6.6 hours; the average is 7.5 hours. We therefore argue that most likely we are viewing W0607+24 close to pole-on. This also explains the lack of variability in the two Spitzer bands, which is otherwise surprising. Based on their Spitzer 3-5$\mu$m monitoring of a large sample of L and T dwarfs, \citet{2015ApJ...799..154M} conclude that spots are present on ``virtually 100\% of L3 -- L9.5 dwarfs" and that their non-detections can be explained by viewing geometry. This would suggest that the probabilities of long periods and $i> 20^\circ $ in Table~\ref{simulation} are overestimated and favors the pole-on scenario.

\begin{deluxetable*}{crrrrrr}
\tabletypesize{\footnotesize}
\tablenum{2} 
\tablecaption{Period/Inclination Posterior Probabilties}
\tablehead{
\colhead{Period (hr)} & 
\colhead{Any $i$} & 
\colhead{$i<5^\circ$} &
\colhead{$5^\circ \le i < 10^\circ$} &
\colhead{$10^\circ \le i < 20^\circ$} &
\colhead{$i \ge 20^\circ$}}
\startdata
Any $P$            & 100\% & 11\% & 26\% & 40\%   & 22\% \\ 
$P<5$               & 28\%   & 7\% & 15\%   & 6\%     & 0\% \\
$5 \le P < 7.5$  & 32\%   & 3\% & 8\% & 21\%   & 0\% \\
$7.5 \le P < 10$ & 20\%  & 1\%   & 2\% & 10\%     & 7\% \\
 $10 \le P < 15$ & 15\%  & 0.3\%   & 1\%   & 4\%     & 10\% \\
$P \ge 15$         &  5\%   & 0\% & $0.1$\% & 0.4\% & 5\% 
\enddata
\label{simulation}
\end{deluxetable*}

\subsection{Implications of Radio Emission}

\begin{figure*}[t]
  \includegraphics[width=\linewidth]{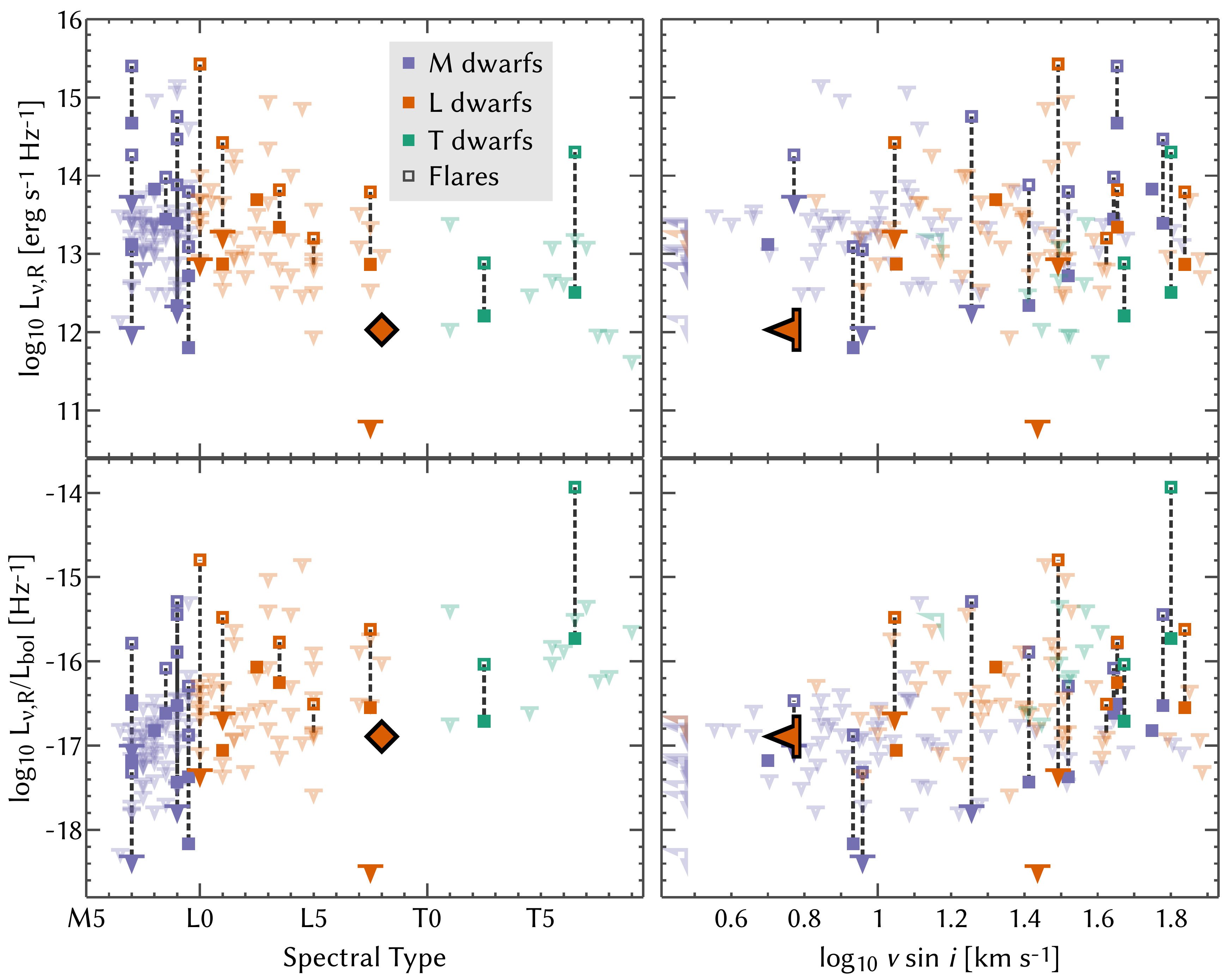}
  \caption{Radio spectral luminosities (\slr) of ultracool dwarfs with
    rotation measurements (Data from \citealt{2014ApJ...785...10C} and references therein, 
    \citealt{2015AJ....150..180B}, \citealt{2016ApJ...818...24K}, \citealt{2016MNRAS.457.1224L})
   In the lower panels these luminosities are scaled
    by \lbol. W0607+24 is highlighted with a large, black-outlined symbol.
    Note that while W0607+24 has a low \vsi, it is not necessarily a slow
    rotator if $\sin i$ is small as we hypothesize. The extremely low upper
    limit at spectral type L7.5 is the nearby (2~pc) binary Luhman~16~AB
    \citep{2015ApJ...805L...3O}. Several more late-L and T~dwarfs have been
    detected in the radio but do not have published rotation rate measurements
    \citep{2016ApJ...818...24K}.}
  \label{f.radiosample}
\end{figure*}

The flux density of our radio detection of W0607+24 corresponds to a spectral
luminosity of $10^{12.0\pm0.2}$ \slumcgs\ and $\slr / \lbol = 10^{-16.9 \pm
  0.2}$~Hz$^{-1}$. Figure~\ref{f.radiosample} places this result in context by
comparing W0607+24 to other ultracool dwarfs with measurements of both radio
emission and rotation. It is an unusual object in several respects: of the
radio-detected L and T dwarfs it has the lowest \slr, it is the only
radio-detected ultracool dwarf for which a measurement of \vsi\ has yielded an
upper limit rather than a detection, and it is the only radio-detected ultracool
dwarf without an H$\alpha$ detection ($\log L_{{\rm H}\alpha}/L_{\rm bol} <3\times10^{-7}$). 
Only the M9.5 dwarf WISE~J072003.20$-$084651.2 has been detected at a lower radio spectral
luminosity \citep{2015AJ....150..180B}. While W0607+24 is clearly one of the
most radio-faint ultracool dwarfs yet detected --- an unsurprising fact given
its proximity --- two effects combine to cause it to appear relatively
radio-bright in the lower-right ($\slr/\lbol$ vs. \vsi) panel of
Figure~\ref{f.radiosample}: its has unusually low values of both \lbol\ and \vsi.
The low \vsi\ of W0607+24 is particularly worthy of note because, as we have
described, it may in fact be due to a nearly pole-on viewing geometry rather
than slow rotation.

W0607+24 is a valuable test case for understanding the mechanisms of radio
emission in ultracool dwarfs. Some of the radio-detected objects emit rapid,
intense bursts with high circular polarization, often found to occur
periodically \citep{Hallinan:2006vn, Hallinan:2008lr, 2009ApJ...695..310B,
  2015ApJ...808..189W} and attributed to coherent emission due to the electron
cyclotron maser instability \citep[ECMI;][]{1979ApJ...230..621W,
  2006A&ARv..13..229T}. Assuming that ECMI bursts propagate perpendicular to
the magnetic field and originate near closely-aligned rotational and magnetic
poles, they may only be detectable from ultracool dwarfs with $i \approx
90\degr$. There is evidence for high inclinations in three of the well-known
ECMI-bursting sources \citep[\object{TVLM 513--46546}, \object{2MASS
    J00361617+1821104}, and \object{LSR J1835+3259};][]{Hallinan:2008lr}. If
the orientation of W0607+24 is truly pole-on, the detection of an ECMI burst
from it would be surprising in this context.

Ultracool dwarfs can also produce steady, broadband emission with weak to
moderate polarization, as we have detected here. This emission is generally
interpreted as being due to the gyrosynchrotron mechanism
\citep[e.g.,][]{Berger:2002fk, 2006ApJ...637..518O, Burgasser:2013qy};
assuming a pole-on orientation, our detection casts further doubt on the
alternative hypothesis that the emission is actually depolarized, steady ECMI
emission \citep{Hallinan:2006vn, Hallinan:2008lr}. Regardless of the
mechanism, there is little insight as to the geometry of the non-flaring
emission region in these objects: its area is likely a few times that of the
(sub)stellar disk \citep[e.g.,][]{Berger:2002fk}, and rotational modulation
suggests that it is not totally axisymmetric
\citep[e.g.,][]{McLean:2011qy}. The relative radio-faintness of W0607+24
may be due to its orientation if the emission is concentrated at the poles as
has been inferred for Algol from VLBI data \citep{1998ApJ...507..371M}.

Meanwhile, if W0607+24 is instead truly a slow rotator rather than orienated
pole-on ($P \gtrsim 32$~hours as per Section~\ref{orientation}), it would be
the slowest-rotating ultracool dwarf with a radio detection, with a rotation
period $\gtrsim$8 times longer than that of \object{NLTT~33370~B}
\citep{2015ApJ...799..192W}, although several radio-active dwarfs have no
rotation measurements available. Assuming a convective turnover timescale
$\tau_c = 70$~days, its Rossby number $\text{Ro} = P / \tau_c \gtrsim 0.02$,
high but not unheard-of within the radio-active ultracool sample
\citep{2014ApJ...785...10C} and still well within the rotation-independent
regime inferred from some geodynamo simulations \citep{2006GeoJI.166...97C}.
Its radio luminosity is just consistent with the slowly-rotating and X-ray
flaring ``group B'' categorization of \citet{2012A&A...537A..94S}, although
the possibility of detecting X-rays from this object seems remote given the
dramatic drop in $L_X$ with temperature \citep{2010ApJ...709..332B,
  2014ApJ...785....9W}.

\section{Conclusions}

Although a typical late-L dwarf in most respects, W0607+24's unusually sharp lines and lack of variability is consistent with a viewing angle near pole-on.  As the nearest known northern hemisphere late-L dwarf, this offers many possibilities for future studies. For example, it could serve as an useful standard for line-broadening and polarization studies. If there is a planetary system aligned with the primary's rotation axis, then we would expect very little radial velocity signal but a strong astrometric signal.  Further follow-up is needed to confirm and better characterize the radio emission.  

The {\it Kepler} and K2 studies of the L dwarfs W1906+40 and W0607+24 demonstrate the value of space-based monitoring with long time baselines. We are monitoring additional late-M and L dwarfs in other K2 campaigns. They are more distant than W0607+24, but because they are warmer, some are brighter and higher signal-to-noise than W0607+24. In addition, most K2 fields are less crowded than Campaign 0. The most promising potential brown dwarf target for K2 is the nearby L5 dwarf 2MASSW J1507476-162738 \citep{2000AJ....119..369R}, one of the brightest and best-studied L dwarfs. It is known to be rapid rotator ($P=2.5$ hr) seen to be spotted in some epochs \citep{2015ApJ...799..154M} but not others \citep{2013MNRAS.428.2824K}. It could be observed in K2 Campaign 15, where we would expect to detect clear signatures of rotational modulation and spot evolution over the three month campaign.

\acknowledgments

This paper includes data collected by the Kepler mission. Funding for the Kepler mission is provided by the NASA Science Mission directorate. The material is based in part upon work supported by NASA under award No. NNX15AV664G. This work is based in part on observations made with the Spitzer Space Telescope, which is operated by the Jet Propulsion Laboratory, California Institute of Technology under a contract with NASA. Support for this work was provided by NASA through an award issued by JPL/Caltech. EB acknowledges support from the National Science Foundation through Grant AST-1008361. 
This work is based in part on observations obtained at the Gemini Observatory, which is operated by the Association of Universities for Research in Astronomy (AURA) under a cooperative agreement with the NSF on behalf of the Gemini partnership: the National Science Foundation (United States), the Science and Technology Facilities Council (United Kingdom), the National Research Council (Canada), CONICYT (Chile), the Australian Research Council (Australia), CNPq (Brazil) and CONICET (Argentina). Some of the data presented herein were obtained at the W.M. Keck Observatory, which is operated as a scientific partnership among the California Institute of Technology, the University of California and the National Aeronautics and Space Administration. The Observatory was made possible by the generous financial support of the W.M. Keck Foundation.  The National Radio Astronomy Observatory is a facility of the National Science Foundation operated under cooperative agreement by Associated Universities, Inc. 

This research has made use of NASA's Astrophysics Data System, the VizieR catalogue access tool, CDS, Strasbourg, France, IRAF, PyQt, and Astropy, a community-developed core Python package for Astronomy \citep{2013A&A...558A..33A}. IRAF is distributed by the National Optical Astronomy Observatory, which is operated by the Association of Universities for Research in Astronomy (AURA) under cooperative agreement with the National Science Foundation. IRAF is distributed by the National Optical Astronomy Observatory, which is operated by the Association of Universities for Research in Astronomy (AURA) under cooperative agreement with the National Science Foundation. This work made use of PyKE \citep{2012ascl.soft08004S}, a software package for the reduction and analysis of Kepler data. This open source software project is developed and distributed by the NASA Kepler Guest Observer Office. This research made use of Montage, funded by the National Aeronautics and Space Administration's Earth Science Technology Office, Computation Technologies Project, under Cooperative Agreement Number NCC5-626 between NASA and the California Institute of Technology. Montage is maintained by the NASA/IPAC Infrared Science Archive. This research made use of APLpy, an open-source plotting package for Python hosted at http://aplpy.github.com 

{\it Facilities:} \facility{EVLA}, \facility{Gemini:Gillett (GMOS)}, \facility{Keck:II (NIRSPEC)}, \facility{Kepler}, \facility{Spitzer (IRAC)}

\bibliographystyle{yahapj}
\bibliography{../astrobib}


\end{document}